\documentstyle[aps,manuscript]{revtex}
\draft
\begin{document}
\title{Neutrinos in a gravitational background: \\
a test for the universality of the gravitational interaction}
\author{H. Casini$^\dagger$, J.C. D'Olivo$^*$, R.  Montemayor$^\dagger$, L.F.
Urrutia$^* $}
\address{\vskip .3cm$^\dagger$Centro At\'omico Bariloche and Instituto
Balseiro\\CNEA and Universidad Nacional de Cuyo\\8400 S.C. de Bariloche,
R\'{\i}o Negro, Argentina\\
\vskip .3cm$^*$ Departamento de F\'\i sica de Altas Energ\'\i as - Instituto
de Ciencias Nucleares\\ Universidad Nacional Aut\'onoma de M\'exico \\
Apartado Postal 70-543\\ 04510 M\'exico, Distrito Federal, M\'exico}
\maketitle

\begin{abstract}
In this work we propose an extended formulation for the interaction
between
neutrinos and gravitational fields. It is based on the parametrized
post-Newtonian aproach, and includes a violation of the universality of the
gravitational interaction which is non diagonal in the weak flavor space. We
find new effects that are not considered in the standard  scenario for
violation of the equivalence principle.  They are of the same order as the
effects produced by the Newtonian potential, but they  are highly directional
dependent and could provide a very clean test of that violation.
Phenomenological consequences are briefly discussed.
\end{abstract}
\pacs{PACS numbers: 04.80.Cc, 14.60.Pq }

\section{Introduction}

Despite the great success of general relativity to explain the gravitational
interaction, this theory has proved to be very difficult to test in detail.
There are several systems where gravitation is important, such as the very
early universe, pulsars, quasars, black holes, and gravitational waves.
However, gravitational fields in astrophysical systems can be considered as
weak, even in the extreme cases of the neighborhood of a neutron star or at
few Schwartzchild radii from a black hole. Usually the gravitational effects
beyond the Newtonian level are very small and too tightly interwoven with
other local physical effects to be clearly observed.

The interest in these effects is twofold. They can shed new light on the
character of the gravitational interaction, and they can provide invaluable
information on some astrophysical systems, such as supernovas and neutron
stars. With respect to the first point, two questions arise naturally:  the
validity of general relativity as a description of the gravitational
interaction, and the universality of this interaction.

In fact, the strongest evidence of the universality of the gravitational
interaction involves electrons, protons and neutrons, that is the members of
the lightest family of matter fields in the standard model. Tests of the weak
equivalence principle for these particles include laboratory experiments of
the E\"{o}tv\"{o}s-type, which measure the gravitational acceleration of
macroscopic bodies. They state that gravity accelerates all macroscopic
objects at the same rate to an accuracy of one part in
$10^{12}$\cite{eotvos}. The experimental limits for the universality between
matter fields and gauge fields are weaker. For example, the supernova SN1987A
gave the opportunity of a direct comparison between the transit time for
photons and neutrinos traversing the same path in a gravitational potential
$\phi ({\bf {r})}$, which leads to limits on the violation of the weak
equivalence principle by massless particles of the order of $\left| \gamma
_{\gamma }-\gamma _{\nu }\right| \lesssim 10^{-3}$, where $\gamma $ is the
PPN (parametrized post-Newtonian formalism) parameter for the scalar
potential\cite{supern}.

It is more difficult to obtain observational evidence on the gravitational
coupling of the heavier families, except for the case of the kaon system
where the bound $\left| \phi \Delta \gamma \right| \leq 2\times
10^{-13}$\cite{kenyon} has been set. Information of this kind can also be
acquired from the propagation of neutrinos in a medium with a gravitational
background. As is well known, ordinary matter affects the neutrino
propagation in a flavor dependent way and, under favourable conditions, large
transformations of one neutrino flavor into another can take place, even for
small mixing between the mass eigenstates. The implications of this mechanism
in astrophysics and cosmology has been extensively examined during the last
years. A similar resonant enhancement could be induced by a non-universality
in the gravitational interaction of the neutrinos, even if they are massless.
Violations of the equivalence principle of the order of $10^{-20}$ cannot be
ruled out in this context, and in fact the observed deficit in the solar and
atmospheric neutrino fluxes have already been interpreted as a positive
signal of this violation\cite{halprin,gasperini}.  An interpretation in terms
of a violation of the Lorentz invariance is also possible\cite{glashow}, but
this could be included within the violation of the equivalence principle
scheme in the case of constant fields\cite {glashleu}. Note that neutrinos
are unequaled as test particles for probing the gravitational field. Because
of the smallness of their interactions, the level of accuracy that can be
achieved with them is several orders of magnitude better than in any other
previous test.

The present work is partially motivated by the above considerations and
develops a general framework for analyzing the possible flavor dependence of
the gravitational interaction. Our approach is a generalization of the one
proposed in Refs.\cite{halprin,gasperini}, where only the effects due to the
scalar gravitational potential $\phi $ was considered. In this way, we find
new contributions to the oscillations that are of the same order of magnitude
than the terms involving the Newtonian potential, and we extend the analysis
to the next PPN order, which includes contributions generated by the angular
momentum of the gravitational source. In contrast with the scalar potential
contribution, this gives place to new highly anisotropic effects, which in
principle could be verified in several astrophysical systems. These new
effects provide a more precise and characteristic signature for a possible
violation of the equivalence principle.

Whereas the integer spin fields can be consistently described in a curved
space-time, the half-integer spin fields need to be defined with reference to
a locally inertial frame at each point of the space- time\cite{weinberg}. If
we are considering several neutrino flavors, we have to extend the usual
construction by defining each gravitational flavor in its own inertial frame.
Furthermore, we can introduce a possible violation of the equivalence
principle and assume that these frames are not necessarily related by Lorentz
transformations, and thus could be physically non-equivalent. At each of the
orthonormal frames the gravitational field is supposed to have the structure
given by the PPN formalism, which provides a general account of the possible
deviations from the Einstein theory.

These assumptions lead us to a generalization of the standard scenario for
the violation of the equivalence principle (VEP). With our present
theoretical understanding it seems that if we want to keep the spinor
structure, the theory does not correspond to a metric one which can be
settled on a consistent basis. More precisely, our approach should be
considered as a phenomenological one, which allows us to search for possible
signatures of flavor-changing effects associated with a violation of the
equivalence principle. In the following we do not consider the dynamics of
the gravitational fields but we still use a manifold as the space-time
framework, i.e., although the metric is not defined we assume that the
space-time is well defined.

\section{Neutrinos in a gravitational background}

Given the frame $V_{\alpha }^{\mu }$ the equation for a (massless) neutrino
is the Dirac equation in a curved space time:
\begin{equation}
\gamma ^{\alpha }V_{\alpha }^{\mu }\,(\partial _{\mu }+i\,\Gamma _{\mu
})\,\Psi =0\,,  \label{2}
\end{equation}
where the connection is
\begin{equation}
\Gamma _{\mu }=-\frac{1}{2}V_{\nu }^{\alpha }\,\nabla _{\mu }V^{\nu \beta
}\sigma _{\alpha \beta }\,,  \label{3}
\end{equation}
with $\sigma _{\alpha \beta }=\frac{i}{2}[\gamma _{\alpha},\gamma_{\beta}]$.
All covariant derivatives are metric, so we are neglecting small torsion
effects\cite{torsion}. Other possibilities for the equation involve generical
couplings with the curvature\cite{ng}, but these terms are highly suppressed
in the usual astrophysical situations by the small values of the gradients of
the gravitational fields as compared to the neutrino momentum.  Explicitly,
the linearized Dirac equation for a static gravitational field
reads\cite{cm}:
\begin{equation}
\left( i\gamma ^{\mu }\partial _{\mu }-\frac{i}{4}\left\{ h_{00},{\bf {
\gamma \cdot \nabla }}\right\} -\frac{i}{4}\left\{ h_{ij},\gamma
^{j}\partial ^{i}\right\} +\frac{i}{2}\left\{ h_{0i},\gamma ^{0}\partial
^{i}\right\} +\frac{1}{2}\gamma ^{0}\epsilon ^{ijk}\partial
^{i}h^{0j}s^{k}\right) \Psi _{\nu }=0\,,  \label{LINDIR}
\end{equation}
where the $h^{\mu \nu }$ fields are defined by $g^{\mu \nu }=\eta ^{\mu\nu
}+h^{\mu \nu }$ and $s^{k}=\frac{1}{4}\epsilon ^{ijk}\sigma ^{ij}$. The
spatial derivatives of the gravitational potentials are proportional to the
inverse of their characteristic variation length, $\partial _{i}h^{\mu \nu
}\propto L^{-1}$. In astrophysical systems the spatial dimensions and the
energy of the neutrinos render $L\gg \lambda _{\nu }$, where $\lambda _{\nu
}\propto p^{-1}$ is the neutrino wavelength. Accordingly, we can neglect the
terms with spatial derivatives of the gravitational fields, including the
spin contributions. This approximation can be justified on a more general
basis by using a geometric optic-like expansion of the Dirac equation in
powers of the parameter $\lambda /L$\cite{anandam}. The main effects of the
neglected terms, which include a chirality transition induced by gravity, are
independent of the equivalence principle violation, and have been already
analyzed in Ref. \cite{cm}. In the above approximation Eq.  (\ref {LINDIR})
reduces to
\begin{equation}
i\partial _{0}\Psi _{\nu }=H\Psi _{\nu }\,,  \label{PHAM}
\end{equation}
with the Hamiltonian given by
\begin{equation}
H=-i\gamma _{0}\gamma _{i}[(1-\frac{1}{2}h^{00})\partial _{i}- \frac{1}{2}
h_{ij}\partial _{j}]-ih_{0i}\partial _{i}\;,  \label{HAM}
\end{equation}
where all metric dependent terms are assumed to be slowly varying functions
of the position.

In an astrophysical scenario we can work within the framework of the PPN
theories\cite{wheeler}. The assumptions for constructing the metric in the
PPN formalism involve virialized sources such that $\frac{M}{R}\sim
{\rm w}^{2}$, where the quantities $M,R$, and $\rm w$ represent estimations
of the order of magnitude of the mass, distance and characteristic (average)
velocity of the source.

The metric is the Minkowskian one plus source dependent perturbations. The
latter have the correct tensorial character and dimensions, falling at least
like $1/R$ at infinity. This metric is in general given by
\begin{eqnarray}
h_{oo} &=&2\gamma ^{\prime }U+{\cal O}({\rm w}^{4})\;, \\
h_{oi} &=&-\frac{7}{2}\Delta _{1}V_{i}-\frac{1}{2}\Delta_{2}W_{i}+(\alpha
_{2}-\frac{1}{2}\alpha _{1}){\rm v}_{i}U-\alpha _{2}{\rm v}_{j}U_{ji}+{\cal
O}({\rm w}^{4}) \;,  \label{PPN} \\
h_{ij} &=&2\gamma U\delta _{ij}+\Gamma U_{ij}\,+{\cal O}({\rm w}^{4})\;,
\end{eqnarray}
where the potentials are
\begin{eqnarray}
U &=&\int \frac{\rho ({\bf r}^{\prime })\;d^{3}r^{\prime }}{\mid {\bf r}-
{\bf r}^{\prime }\mid }\;,  \label{UVUW} \\
U_{ij} &=&\int \frac{\rho ({\bf r}^{\prime })(r_{i}-r_{i}^{\prime
})(r_{j}-r_{j}^{\prime })\;d^{3}r^{\prime }}{\mid {\bf r}-{\bf r}^{\prime
}\mid ^{3}}\;,   \\
V_{j} &=&\int \frac{\rho ({\bf r}^{\prime }){\rm w}_{j}({\bf r}^{\prime
})\;d^{3}r^{\prime }}{\mid {\bf r}-{\bf r}^{\prime }\mid }\;, \\
W_{j} &=&\int \frac{\rho ({\bf r}^{\prime })({\bf w}({\bf r}^{\prime })\cdot
({\bf r}-{\bf r}^{\prime }))(r_{j}-r_{j}^{\prime })\;d^{3}r^{\prime }}{\mid
{\bf r}-{\bf r}^{\prime }\mid ^{3}}\;.\label{UVUW2}
\end{eqnarray}
Here $\rho ({\bf r})$ is the density of mass and ${\bf w}({\bf r})$ is the
velocity of the source of the gravitational field. We are using a system of
unities where $G=\hbar =c=1$. We keep each term of the same order in the PPN
expansion because we are interested in ultrarelativistic neutrinos.

In the particular case of a very confined and distant source,  the
expressions in Eqs. (\ref{UVUW})-(10) can be approximated as follows
\begin{eqnarray}
U &\approx &\frac{M}{R}+O\left( \frac{1}{R^{2}}\right)
\;,\;\;\;\;\;\;\;\;\;\;\;\;U_{ij}\approx \frac{X_{i}X_{j}}{R^{2}}U
+O\left(\frac{1}{R^{2}}\right) \;,  \\
V_{j} &\approx &{\rm w}_{j}U+O\left( \frac{1}{R^{2}}\right)
\;,\;\;\;\;\;\;\;\;\;\;\;W_{j}\,\approx \frac{{\rm w}_{i}X_{i}X_{j}}{R^{2}}
U+O\left( \frac{1}{R^{2}}\right) \;,
\end{eqnarray}
where $X_{i}$ are the components of ${\bf R}$.

Up to order ${\rm w}^{3}$ the adimensional parameters of the expansion are
$\gamma $, $\gamma ^{\prime }$, $\Delta _{1}$, $\Delta _{2}$, $\Gamma $,
${\bf {v}}$ , $\alpha _{1}$ and $\alpha _{2}$. In Einstein gravity we have
$\alpha _{1}=\,\alpha _{2}=\Gamma =0\,,\gamma =\gamma ^{\prime }=\Delta
_{1}=\Delta _{2}=1$\thinspace and ${\bf {v}}$ is irrelevant. The parameters
$\alpha _{1}$ and $\alpha _{2}$ are null if the theory is Lorentz covariant,
but if there is a preferred reference frame, characterized by a velocity
${\bf {v,}}$ they should be non null. In general the characteristic frame
velocity can be a gravitational flavor dependent quantity. The parameter
$\alpha _{1}$ can be fixed to be $7\Delta _{1}+\Delta _{2}-4\gamma -4\gamma
^{\prime }$, while $\alpha _{2}$ is an independent parameter up to this
order.

The dispersion relation implied by the Hamiltonian in Eq.(\ref{HAM}) is
\begin{equation}
(1-h^{00}){\bf p}^{2}-E^{2}+2Eh_{0i}p_{i}-h_{ij}p_{i}p_{j}=0\;, \label{DR}
\end{equation}
where $E$ is the energy eigenvalue and we keep only first order terms in the
metric $h_{\mu \nu }$.

\section{Neutrino oscillations from a non-universal gravitational interaction
}

The purpose of our work is to state a meaningful phenomenological basis to
discuss a possible violation of the equivalence principle. The breakdown of
the universality of the gravitational interaction raises the question of the
complete consistency of the underlying theory, but we will put aside this
issue. Here, we are concerned only with a model that describes the different
phase shifts associated to a possible flavor dependence of the gravitational
interaction, its consequences, and the possibilities of its detection in an
astrophysical framework. We assume that the metric associated with each
family are very close to each other and to the Einstein theory predictions.
The PPN coefficients depend on the flavor and are assumed to be diagonal
matrices in the gravitational flavor basis. The more general case where these
coefficients are arbitrary matrices would imply that there exist no local
inertial frames where the neutrinos can be defined, and therefore their
spinorial nature would not be clear. We do not consider this situation.

Following this approach, each of the $n$ neutrino flavors is defined in a
different orthonormal frame $\{V_{a\,\mu }^{\alpha }\,\},\;a=1,...,n$, which
form non-equivalent bases for the tangent space.  Accordingly, we write the
metrics as
\begin{equation}
\eta _{\alpha \beta }\,V_{a\,\mu }^{\alpha }V_{a\,\nu }^{\beta}=g_{\mu \nu
}^{a}\;\,.  \label{1}
\end{equation}

From Eq. (\ref{DR}), the dispersion relation for each neutrino gravitational
flavor can be approximated by
\begin{equation}
E^{a}=p\left( 1+h_{oi}^{a}\,\hat{p}_{i}\,-\Gamma
^{a}\,U_{ij}\hat{p}_{i}\hat{
p}_{j}-(\gamma ^{\prime a}+\gamma ^{a})U\right) \,.
\label{dispersion}
\end{equation}
In the PPN approximation the coordinate system is generally fixed to give
$\Gamma =0$ and $\gamma^{\prime}=1$, (where the last equality is equivalent
to the definition of the Newton constant,) but in our case such a coordinate
fixing could be done only for one metric at the expense of the others.
Therefore, we are setting these parameters at the usual values for the first
gravitational family, while the others are left as free parameters.  This is
an important point as we will see in what follows.

If the parameters are family dependent, then distinct neutrinos will undergo
different phase shifts when passing through the same sector of the space.
The phase shift differences become observable when the particle basis that
diagonalize the weak and the gravitational interaction are not the same. In
this context, the effects of a universality violation at the level of the
scalar potential $U$ has been already considered in Ref.  \cite{halprin},
where it is shown that such a violation leads to neutrino oscillations. The
factor $(1+\gamma ^{a})$ of their model\ should be replaced by $(\gamma
^{\prime a}+\gamma^{a})$ in our case. We extend this model by including the
effects of the $U_{ij}$ potentials that are of the same order of magnitude as
$U$, together with the PPN structure expanded up to $w^{3}$, with the
corresponding generalized universality violations. In this way, we are taking
into account not only the coupling of the neutrinos to the mass of the source
of the gravitational field, but also the coupling to its quadrupolar
distribution and its angular momentum. As we will show, these last
interactions produce highly directional and very characteristic effects.

In what follows we examine the neutrino oscillations induced by a
non-universal gravitational coupling. To find out the main features of this
phenomena we will consider two neutrino flavors. We assume that the
gravitational flavor basis is related to the electroweak basis through a
unitary transformation ${\cal U}$, characterized by a mixing angle $\theta
_{g}$:
\begin{equation}
\left(
\begin{tabular}{l}
$\nu _{1}$ \\
$\nu _{2}$
\end{tabular}
\right) ={\cal U^{\dagger }}\left(
\begin{tabular}{l}
$\nu _{e}$ \\
$\nu _{\mu }$
\end{tabular}
\right) \equiv \left(
\begin{tabular}{ll}
cos$\theta _{g}$ & -$\sin \theta _{g}$ \\
sin $\theta _{g}$ & $\cos \theta _{g}$
\end{tabular}
\right) \left(
\begin{tabular}{l}
$\nu _{e}$ \\
$\nu _{\mu }$
\end{tabular}
\right) \;.  \label{CFB}
\end{equation}

The equation for the neutrino evolution is
\begin{equation}
i\frac{d}{dt}\left(
\begin{tabular}{l}
$\nu _{e}$ \\
$\nu _{\mu }$
\end{tabular}
\right) ={\cal U}{\cal HU^{\dagger }}\left(
\begin{tabular}{l}
$\nu _{e}$ \\
$\nu _{\mu }$
\end{tabular}
\right) \;,
\end{equation}
where ${\cal H}$ is a diagonal matrix in the gravitational flavor basis,
whose eigenvalues are given by Eq. (\ref{dispersion}). After discarding an
irrelevant overall phase, we have
\begin{equation}
i\frac{d}{dt}\left(
\begin{tabular}{l}
$\nu _{e}$ \\
$\nu _{\mu }$
\end{tabular}
\right) =\frac{\Delta _{0}}{2}\left(
\begin{tabular}{ll}
-cos$2\theta _{g}$ & $\sin 2\theta _{g}$ \\
\ $\sin 2\theta _{g}\,$ & cos$2\theta _{g}$
\end{tabular}
\right) \left(
\begin{tabular}{l}
$\nu _{e}$ \\
$\nu _{\mu }$
\end{tabular}
\right) \;,  \label{alfa}
\end{equation}
with
\begin{equation}  \label{DELTA0}
\Delta _{0} = E_2-E_1=E\left[ \delta h_{0i}\,\hat{p}_{i}\,-\delta \Gamma
\,U_{ij}\hat{p}_{i}\hat{p}_{j}-(\delta \gamma ^{\prime }+\delta \gamma )U
\right],
\end{equation}
where $E=p$ is the neutrino beam energy, and
\begin{equation}
\delta \gamma = \gamma ^{2}-\gamma ^{1},\quad \delta \gamma ^{\prime
}=\gamma ^{\prime 2}-\gamma ^{\prime 1}, \quad \delta \Gamma =\Gamma
^{2}-\Gamma ^{1},
\end{equation}
\begin{equation}
\delta h_{0i}=h_{0i}^{2}-h_{0i}^{1}\simeq -\frac{7}{2}\delta \Delta
_{1}V_{i}-\frac{1}{2}\delta \Delta _{2}W_{i}+(\delta \alpha _{2}- \frac{1}{2}
\delta \alpha _{1}){\rm v}_{i}U-\delta \alpha _{2}{\rm v}_{j}U_{ji}\;\,.
\end{equation}

For a constant gravitational field the survival probability is
\begin{equation}
P(\nu _{e}\rightarrow \nu _{e})=1-\sin ^{2}\left( 2\theta _{g}\right) \,\sin
^{2}\left( \frac{\pi L}{\lambda _{g}}\right) \;,  \label{SP}
\end{equation}
where $L=t-t_{0}$ is the distance travelled by the neutrino from the
production point. This clearly shows that oscillations will appear whenever
there exists a non null mixing angle induced by flavor dependent
gravitational interactions. These oscillations have a characteristic length
given by
\begin{equation}
\lambda _{g}=\frac{2\pi }{|\Delta _{0}|}\,\;.  \label{lg}
\end{equation}

In contrast with vacuum oscillations induced by a mass difference, where
$\lambda_{m}=\frac{4\pi E}{\delta m^{2}}$ is proportional to the energy, the
effect we are considering has oscillation lengths proportional to $E^{-1}$,
which makes this phenomena suitable to be observed in the case of high energy
neutrinos. Note that even though the overall sign of the gravitational
potential is irrelevant for the oscillations, the relative signs among the
parameter differences are very significative. The Eq. (\ref {SP}) leads to
the following averaged survival probability
\begin{equation}
<P(\nu _{e}\rightarrow \nu _{e})>=\frac{1}{2}\left( 1+{\rm \cos }^{2}2\theta
_{g}\right) \;.  \label{ASPV}
\end{equation}

As is well known, neutrino oscillations in matter are qualitatively different
from the oscillations in the vacuum. This is because the interaction of the
neutrinos with matter modify their dispersion relations.  Neutral current
interactions are flavor diagonal and can be ignored, as long as we do not
consider sterile neutrinos and neutrinos are not part of the medium. In
general, this will not be true for the charged current interactions. The
forward scattering amplitude is not flavor diagonal in this case, and depends
on the leptonic content of the matter. The above gives place to important
consequences such as the MSW effect. In general, we expect that a non
universal gravity will also affect the electroweak Lagrangian, introducing a
number of unknown coefficients, but the combined effect should be of the
order $U\,G_{F}\,$ and therefore is highly suppressed.

If electrons are the only leptons that are present, then the matrix
\begin{equation}
\frac{b_{e}(t)}{2}\left(
\begin{tabular}{ll}
$1$ & $\ \ 0$ \\
$0$ & $-1$
\end{tabular}
\right) \;,
\end{equation}
has to be added to the second term of Eq. (\ref{alfa}). Here, $b_{e}(t)=
\sqrt{2}G_{F}N_{e}(t)$, with $N_{e}(t)$ denoting the electron density. The
resulting Hamiltonian ${\cal H}(t)$ can be diagonalized at every moment by
introducing the instantaneous flavor basis, defined in an analogous way  to
Eq. (\ref{CFB}), with $\theta _{g}\rightarrow \theta _{m}(t)$ and
\begin{equation}
{\rm \sin }2\theta _{m}(t)=\frac{\Delta _{0}\ {\rm \sin }2\theta _{g}}{\sqrt{
(\Delta _{0}\ {\rm \cos }2\theta _{g}-b_{e}(t))^{2}+(\Delta _{0}\ {\rm \sin }
2\theta _{g})^{2}}}\;.  \label{THETAM}
\end{equation}

In the adiabatic approximation, the average survival probability is given by
the following formula
\begin{equation}
<P(\nu _{e}\rightarrow \nu _{e})>=\frac{1}{2}\left( 1+{\rm \cos }2\theta
_{g}\ {\rm \cos }2\theta _{m}(t_{0})\right) \;.  \label{ASPM}
\end{equation}
which reduces to Eq. (\ref{ASPV}) when $b_{e}(t)=0$, so that $\theta
_{m}(t)=\theta _{g}$. The use of the adiabatic approximation is justified
whenever
\begin{equation}
\frac{1}{N_{e}(t_{R})}\left| \frac{dN_{e}(t)}{dt}\right| _{t_{R}}<<\,|\Delta
_{0}|\,\frac{{\rm \sin }^{2}2\theta _{g}}{{\rm \cos }2\theta _{g}}\;.
\label{CAA}
\end{equation}

There also exists the possibility of a resonant conversion when the diagonal
elements of the full Hamiltonian vanish, i.e. when \begin{equation}
\sqrt{2}G_{F}N_{e}(t_{R})=\Delta _{0}\,\cos 2\theta _{g}\,.  \label{resconv}
\end{equation} This mechanism can totally change the flavor, independently of
the value of the mixing angle $\theta _{g}$, but its efficiency depends on
the adiabaticity of the process.

\section{Phenomenological effects}

There is an inherent uncertainty in the potentials in Eqs. (\ref{UVUW})-(10),
because arbitrary constants can always be added without changing the physics,
as far as the effects associated with a violation of the equivalence
principle are not involved. In Einstein theory these constants can be
eliminated by a coordinate transformation. Instead, in the PPN expansion the
coordinate system is fixed, so possible uncertainties arise from the very
distant unknown mass distributions. We will restrict ourselves to the most
important known near sources for the potentials, leaving aside the problem of
the very distant ones. In any case, the distant sources would only produce
significant isotropical effects, and thus could only affect the definition of
$U$ and the diagonal part of $U_{ij}$ (which in the formula for $\lambda
_{g}$ can be absorbed into $U$).

\subsection{Solar neutrinos}

The sun is subject to a gravitational field that has several sources.  The
main sources are: our galaxy, the Virgo Cluster, and the Great Attractor.
The more important contribution to the potential $U$ comes from the Great
Attractor gravitational field, with small perturbations due to galactic
clusters, and the galactic and solar fields. In consequence, this potential
can be approximated by a constant of the order of $10^{-5}$\cite
{kenyon,Lynden}. The effect of this potential regarding a possible
universality violation has been already analyzed in Ref.  \cite{halprin}.
Resonant flavor oscillations are consistent with the observed deficit of
solar and atmospheric neutrinos for $U\delta (\gamma +\gamma ^{\prime
})\simeq 10^{-22}$. The effect is isotropic and to account for the more
recent data requires a three-neutrino mixing scheme\cite{halleung}.

In the PPN approximation the contributions of the $U_{ji}$ potential are in
principle considered of the same magnitude as $U$. In the case of a distant
source in the $z$ direction, we have $U_{zz}\sim U$. However, the components
$U_{xx}$, $U_{yy}$ and $U_{xy}$ are proportional to $\left( \Delta \theta
\right) ^{2}U$, where $\Delta \theta $ is the angular size of the source,
while $U_{xz}$ and $U_{yz}$ are of the order of $\Delta \theta \,U$ (in fact,
they are proportional to the center of mass distance to the $z$ axis).
Considering that the Great Attractor is a rather extended object with an
angular size of the order of $10^{-1}$\cite{granatrac}, we see that in the
case of the sun there are only three relevant types of $U_{ji}$
contributions: those coming from our galaxy, which are of the order of
$10^{-6}$, a longitudinal component from the Great Attractor, of order of
$U_{zz}\simeq U\simeq 10^{-5}$, and  transverse-longitudinal components also
produced by the Great Attractor, of the same order as the galactic
contributions, $U_{xz}\simeq U_{yz}\simeq 10^{-6}$. The contributions of
$V_{i}$ and $W_{i}$ are roughly proportional to source velocity ${\bf w}$
times $U$, which is of the order of $(10^{-3}-10^{-2})U$. Their directional
effect has a dipolar structure and can be assumed to be one or two orders of
magnitude smaller than the dominant ones.

Therefore, a violation of the equivalence principle would be characterized
by three main effects manifested as flavor oscillations: an isotropic effect
($U\simeq 10^{-5}$), and two additional anisotropic effects ($U_{zz}\simeq
10^{-5}$, $U_{ji}\simeq 10^{-6}$). If we assume that the differences of the
PPN parameters due to the flavor dependence are all of the same order, the
most significant directional effect is given by a quadrupolar contribution
due to $U_{zz}$. This effect could be of the order of the dipolar one
originated by the elliptical orbit of the Earth, but the latter only depends
on the eccentricity of the orbit (perigee: RA=18:48, DEC=-23:27 in
equatorial coordinates), whereas the gravitational one depends on the energy
of the neutrinos and their direction with respect to the Great Attractor. The
approximate position of the Great Attractor center in galactic coordinates
is $l=325^{o},\;b=-7^{o},\;v=4882\;km\;s^{-1}$, or in equatorial coordinates
is RA=$a$=16:10, DEC=$d$=-$60^o10^\prime$. In ecliptic coordinates the 
aphelion position is RA=18:48 ($282^{o}$), DEC=$0^{o}$, whereas the Great 
Attractor center is at RA=$A$=16:52 ($253^{o}$), DEC=$D$=$-38^{o}17^{\prime 
}$. Both axis differ in approximately $30^{o}$ in RA, and therefore the 
effects could be discriminated. Taking into account the dominant 
contributions, due to $U$ and $U_{zz}$, and assuming $U\simeq U_{zz}$, the 
neutrino wavelength in vacuum becomes:
\begin{equation}
\lambda _{g}=\frac{2\pi }{E\,U\,\left| \delta \Gamma \cos ^{2}D\cos
^{2}\left( \alpha -A\right) +(\delta \gamma ^{\prime }+\delta \gamma
)\right| }\,\;,
\end{equation}
where $\alpha $ is the  right ascension of the sun in ecliptic coordinates
at a given time.

An easily visible consequence of the gravitational contribution is a
breaking of the reflection symmetry of the neutrino flux with respect to the
aphelion-perihelion axis of the Earth orbit. This symmetry is characteristic
of the scenarios which do not consider the violation of the universality of
the gravitational interaction, with the only exception given by a possible
interaction between the solar magnetic field and the neutrino magnetic
moment\cite{akhmedov}. In general, the scenarios usually considered would
yield different neutrino fluxes for Earth positions separated by six months.
Otherwise, the gravitational contribution has the same sign for these two
positions.

\subsection{Atmospheric neutrinos}

The dominant contributions in this case are the same as those already
considered in the previous section. The main difference is originated by the
Earth rotation which gives place to diurnal neutrino flux variation. This
situation can be described more appropriately by means of azimuthal
coordinates\cite{sidgwick}. In terms of these coordinates the neutrino
direction can be written:
\begin{equation}
\hat{p}_{\nu }=(\sin \theta _{\nu }\cos z_{\nu },\sin \theta _{\nu }\sin
z_{\nu },\cos \theta _{\nu })\;,
\end{equation}
where $\theta _{\nu }$ is the zenithal distance and $z_{\nu }$ is the
azimuthal angle of the incident neutrinos. Similarly, for the Great
Attractor position we have:
\begin{equation}
\hat{d}_{GA}=(\cos \varphi \sin d -\sin \varphi \cos d \cos \tau
,\cos d \sin \tau ,\sin d \sin \varphi +\cos d \cos \varphi
\cos \tau )\;,
\end{equation}
with $d$ being the Great Attractor declination angle and $\varphi $
the observatory latitude. The parameter ${\tau }$ is $a-t_{s}$, where $a$
is the right ascension of the Great Attractor and $t_{s}$ is the sidereal
time. According to Eqs. (\ref{DELTA0}) and (\ref{lg}) the oscillation
wavelength depends on $\left( \hat{p}_{\nu }.\hat{d}_{GA}\right) ^{2}$ . If
there were no violation in the universality of the gravitational
interaction, then the neutrino flux would only depend on the zenithal
distance $\theta _{\nu }$. Here we have an additional dependence on the
azimuthal angle $z_{\nu }$ and the time $\tau $, which implies a diurnal
variation of the flux. For instance, if we focus our attention on zenithal
neutrinos ($\theta_{\nu }=0$), for the Kamiokande site ($\varphi=36.5^{o}$)
we have:
\begin{equation}
\lambda _{g}=\frac{2\pi }{E\,U\left| \left( -.5+.4\cos \tau\right)
^{2}\delta \Gamma +(\delta \gamma ^{\prime }+\delta \gamma )\right| }\;.
\end{equation}
This is the effect at a given time $\tau$ on the vacuum wavelength. It can
be traced on with the $\theta _{\nu }$ dependence of the total flux integrated
over $\tau$ and $z_{\nu} $.

\subsection{Neutron star kicks and rotation}

Resonant neutrino oscillations influenced by the magnetic field in the early
stage of a neutron star have been proposed as a possible mechanism to
explain the observed proper motion of pulsars\cite{segre}. But to do this
excessively strong magnetic fields are required\cite{janka}. Furthermore,
the origin of the high angular velocity of the pulsars is also not completely
clear\cite{spruit}. In the present discussion, the vectorial potential
$h_{0i}$ manifests itself as an anisotropic perturbation in the shape of the
resonant surface that is located between the neutrinospheres of the
different flavors. This anisotropy could be the source of both the
angular and the traslational accelerations, and thus could simultaneously
give an explanation for the observed spins and proper motions.

The potential $h_{0i}$ contains two main contributions, both produced by the
star. One of them, given by $(\delta \alpha _{2}-\frac{1}{2}\delta \alpha
_{1}){\rm v}_{i}U-\delta \alpha _{2}{\rm v}_{j}U_{ji}\,$, is relevant when a
preferred frame exists, and therefore the coefficients $\alpha$ are non
null. Their action is analogous to the one produced by a strong magnetic
field\cite {segre,magnetic}, which can generate a translational kick in the
movement of the star during neutrino emission. Taking reasonable values for
the potential and a frame velocity of the order $10^{-3}$, the resulting
gravitational effect has the correct magnitude to explain the observed kicks.
The other contribution to $h_{oi}$ originates from $V_{i}$ and is caused by
the star rotation. It induces an angular acceleration during neutrino
emission because the resonant surface depends on the neutrino angular
momentum. The sign of this acceleration is determined by the relative sign
between $(\delta \gamma ^{\prime }+\delta \gamma )$ and $\delta \Delta _{1}$.

Since the gravitational fields in the interior of a neutron star are
relatively high, $U\lesssim 1$, the PPN approximation is not very good in
this case. A more complete treatment should include higher order terms in the
expansion, and hence more parameters.

\section{Conclusions}

In this work we have formulated a generalized scenario for neutrinos
propagating in the presence of gravity. To incorporate the violation of the
equivalence principle each neutrino flavor has been defined in a different
orthonormal frame at each point of space-time. This is a natural prescription
if we are giving up the metric but still want to retain the manifold
structure. The violation is parametrized by a generalized PPN expansion.

In this way we have developed a VEP scenario for neutrino oscillations,
which leads to several new effects. The most relevant is the one due to the
potential $U_{ji}$, which can be of the same order of magnitude as the
Newtonian contribution. In the PPN metric theories this potential is
irrelevant because it can be set to zero by a coordinate fixing. This is not
the case in the present approach, where there is more than one metric. In
addition, other potentials could be relevant for the early stages of neutron
stars, giving place to angular or translational accelerations.

An interesting feature of the effects we have discussed is that they have a
very characteristic directional dependence. These effects should be present
in the solar and atmospheric neutrino fluxes if their anomalies are related
to a VEP scenario. In such a way, these phenomena could provide relevant
information on this violation. Accordingly, the solar neutrino flux must
change along the orbit of the Earth in a way that clearly differs in the
angular and energy dependence from the geometrical effect due to the orbit
eccentricity and its consequences on the ordinary vacuum oscillations. In
contrast to the mass-mixing scenario, the seasonal effect we consider could
also appear in the case of resonant transitions in the sun, due to the
variation of the position of the resonance region. In the case of atmospheric
neutrinos, a detailed analysis of the daily and zenithal angle dependence of
the flux, could reveal the effects we are interested in.  Further analysis
along these lines are in progress.

\section{Acknowledgments}

This research has been partially supported by CONICET, Argentina, and
CONACYT, M\'exico.

\end{document}